\documentstyle[twoside,fleqn,espcrc2,macros,epsfig]{article}

\newcommand{\AmS}{{\protect\the\textfont2
  A\kern-.1667em\lower.5ex\hbox{M}\kern-.125emS}}

\title{
{
\vspace{-3.0cm} \normalsize \hfill
\parbox{30mm}{DESY 99-152\\September 1999}
}\\[15mm]
String breaking and lines of constant physics in the
SU(2) Higgs model\thanks{
Talk given at 17th International Symposium on Lattice
Field Theory (LATTICE 99), Pisa, Italy, June 29 - July 3, 1999.}
}

\author{F. Knechtli\\[0.2cm]
        DESY, Platanenallee 6, D-15738 Zeuthen, Germany}
       
\begin{document}

\begin{abstract}
We present results for the ground state and first excited state
static potentials in the confinement ``phase'' of the SU(2) Higgs
model. String breaking and the crossing of the energy levels are
clearly visible. We address the question of the cut-off effects in our
results and observe a remarkable scaling of the static potentials.
\end{abstract}

\maketitle

\section{Introduction}

String breaking, the flattening of the static potential,
has been observed by computing on the lattice
the static (fundamental) potential in the three- and four-dimensional
SU(2) Higgs model and the static adjoint potential in the
three-dimensional SU(2) Yang-Mills theory.
In QCD, recent attempts with two flavors of dynamical quarks 
failed to observe string breaking \cite{schilling}.
The problems in the extraction of the static
potential in QCD arise from the poor overlap of the ``string states'',
described by Wilson loops, with the ground state
at large separations of the static quarks. The latter is better described
in terms of two static-light mesons, which are bound states of a
static quark and the light dynamical quark field. To overcome this
problem one has to extract the static potential from a matrix
correlation function in which ``string'' and ``two-meson states''
enter. This basic point was already noted in \cite{adjpot:su2michael}.

We compute the potential between static charges in the fundamental
representation of the gauge group (static quarks)
in the four-dimensional SU(2) Higgs
model in the confinement ``phase'', where it resembles QCD. 
We improve our first study \cite{4dhiggs} with a better lattice
resolution and also determine
the first excited state static potential.
Furthermore, we address the question of finding lines of constant
physics in order to study the scaling behavior.
For a detailed exposition of our results, we refer to \cite{phd}.

\section{String breaking}

We construct a matrix correlation function $C_{ij}(r,t)$
\cite{4dhiggs} using a
basis of states $|i\rangle$ that contains string-type states,
described by (smeared) Wilson lines,
and meson-type states, described by (smeared) Higgs fields.
The static potentials $V_{\alpha}(r)\;(\alpha=0,1,2,...)$
are extracted from $C_{ij}(r,t)$ using the variational method proposed in
\cite{phaseshifts:LW}.
Because of divergent (like $\frac{1}{a}$) self-energy contributions of
the static charges, the static potentials $V_{\alpha}(r)$ need to be
renormalised. This can be done by considering
\bes\label{renpot}
 \rnod\,[V_{\alpha}(r)-2\mu] \,.
\ees
The energy $\mu$ is the mass of a static-light meson, bound state of a
static charge and the dynamical Higgs field.
The multiplication with the
reference scale $\rnod$ \cite{pot:r0}
makes the physical quantity dimensionless. In
\fig{f_invariant_lcp} (circles), we show the
ground state and first excited state static potentials (\ref{renpot})
as functions of the distance $r/\rnod$ for the bare parameter set 
$\beta=2.4$, $\kappa=0.2759$ and $\lambda=0.7$
(we adopt the conventional notation of \cite{Higgs:Montvay1}).
The ground state potential shows an approximate
linear rise at small distances: around distance
\bes\label{rb}
 \rb & \approx & 1.9\,\rnod
\ees
the potential flattens. The string breaks.
As expected, for large distances the potential approaches
the asymptotic value $2\mu$. The first excited potential comes
very close to the ground state potential around $\rb$ and rises linearly at
larger distances.
\begin{figure}[tb]
\hspace{0cm}
\vspace{-1.0cm}
\centerline{\psfig{file=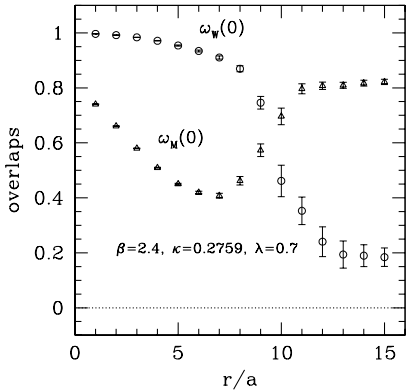,width=6.0cm}}
\vspace{-.5cm}
\caption{{\small Here, we show the overlaps of the string-type 
 (circles) and meson-type (triangles) states
 with the ground state of the Hamiltonian.} 
 \label{f_overlaps0}}
\end{figure}
\begin{figure}[tb]
\hspace{0cm}
\vspace{-1.0cm}
\centerline{\psfig{file=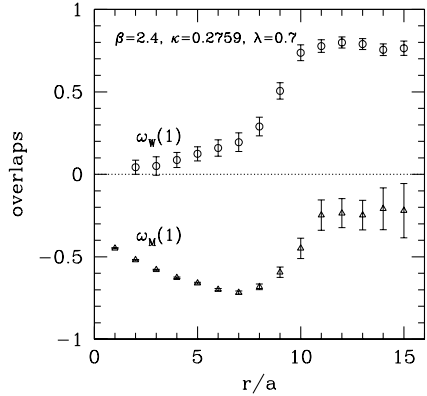,width=6.0cm}}
\vspace{-.5cm}
\caption{{\small Here, we show the overlaps of the string-type
 (circles) and meson-type (triangles) states
 with the first excited eigenstate of the Hamiltonian.} 
 \label{f_overlaps1}}
\end{figure}

We would like to get an insight into the interplay between string-type
and meson-type states.
We consider the
diagonal sub-blocks of the matrix correlation function corresponding
to string-type and to meson-type states separately and
determine approximate ground state wave functions $v_0^{\rmW}$
for the
string-type states and $v_0^{\rmM}$ for the meson-type
states. We then construct a projected matrix
correlation function
\bes\label{projcorrwm}
 \Omega_{kl}(t) & = & v_{0,i}^kC_{ij}(t)v_{0,j}^l \quad
 (k,l=\rmW,\rmM) \nonumber \\
 & = & \sum_{\alpha} \omega_l(\alpha)\omega_k(\alpha)
 \rme^{-tV_{\alpha}(r)} \,,
\ees
where $\alpha$ labels the true eigenstates of the
Hamiltonian and the real coefficients $\omega_k(\alpha)$
are the overlaps of the string-type ($k=\rmW$) and meson-type
($k=\rmM$) states with the eigenstates of the Hamiltonian.
The overlaps with the ground state are shown in \fig{f_overlaps0} and
with the first excited eigenstate in \fig{f_overlaps1} (we choose the
sign conventions $\omega_{\rmW}(0)>0$ and $\omega_{\rmW}(1)>0$).
In the string breaking region around $r/a=9-10$, the
overlaps of the string-type and meson-type states have similar
magnitude, both in \fig{f_overlaps0} and \fig{f_overlaps1}.
The scenario of string breaking as a level crossing
phenomenon \cite{drummond} is confirmed.
We would like to point out that the overlaps represented in
\fig{f_overlaps0} and \fig{f_overlaps1} are not quantities which
have a continuum limit.

\section{Lines of constant physics}
\begin{figure}[tb]
\hspace{0cm}
\vspace{-1.0cm}
\centerline{\psfig{file=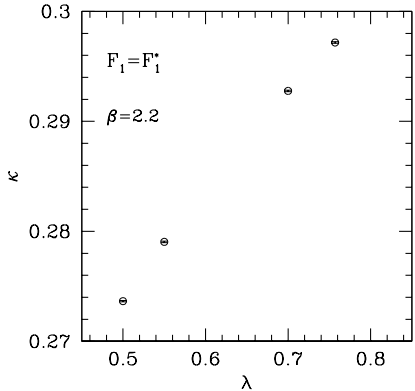,width=6.0cm}}
\vspace{-.5cm}
\caption{{\small Here, we show the matching of $\kappa$ at $\beta=2.2$
 using the matching condition \eq{matching1}.}
 \label{f_lcpkl}}
\end{figure}
\begin{figure}[tb]
\hspace{0cm}
\vspace{-1.0cm}
\centerline{\psfig{file=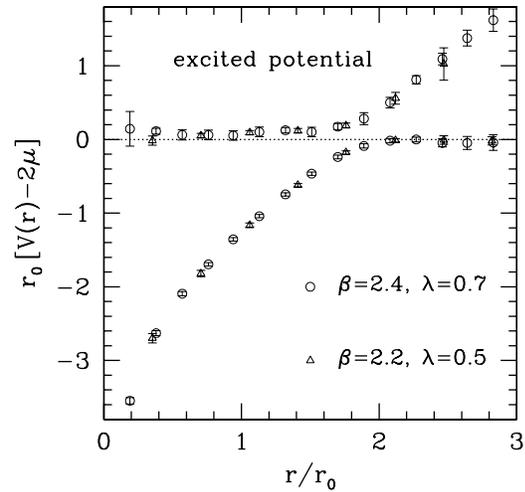,width=7.5cm}}
\vspace{-.5cm}
\caption{{\small Here, we show the scaling of the renormalised ground
 state and first excited state static potentials. The value of
 $\kappa$ is matched to keep $F_1=F_1^*$.} 
 \label{f_invariant_lcp}}
\end{figure}

In order to investigate the presence of lattice artifacts in
dimensionless physical quantities such as (\ref{renpot}),
we must vary the lattice spacing $a$ (by changing $\beta$) and
tune the bare parameters $\kappa$ and $\lambda$ to keep
two dimensionless physical quantities $F_1$ and $F_2$ constant. 
This procedure corresponds to the renormalisation of $\kappa$ and
$\lambda$ and defines in the parameter space a so called line of
constant physics. A sensible choice would be to
find $F_1$ strongly dependent on $\kappa$ and $F_2$ strongly dependent on
$\lambda$.

We consider the quantity
$F_1=\rnod\,[2\mu-V_0(\rnod)]$: it
mainly depends on the value of the mass of the dynamical Higgs field 
which is in turn determined by the choice of $\kappa$.
At $\beta=2.4$ we find
\bes\label{matching1}
  F_1 \; = \; \rnod\,[2\mu-V_0(\rnod)] \; = \; F_1^*\equiv1.26 \,.
\ees
For different values of $\lambda$, we match the
parameter $\kappa$ at $\beta=2.2$ using the condition \eq{matching1}.
The results are shown in \fig{f_lcpkl}.
As a by-product of these simulations we obtain that
\bes\label{aratio}
 {a(\beta=2.2) \over a(\beta=2.4)} & \approx & 1.9 \, .
\ees

We studied several candidates \cite{phd} for a quantity $F_2$ which is
sensitive to a variation of the parameter $\lambda$ once the parameter
$\kappa$ is matched using the condition \eq{matching1}. We cannot draw
a definitive conclusion yet. Our results support earlier observations
\cite{Higgs:Montvay1} that the physics in the confinement ``phase'' of
the SU(2) Higgs model is weakly dependent on the parameter
$\lambda$. We use this as an assumption and study the
scaling behavior of the static potentials between
$\beta=2.2$ and $\beta=2.4$ by using only the matching condition for
$F_1$ \eq{matching1}.

In \fig{f_invariant_lcp}, we compare the results for the
ground state and first excited state static potentials.
They show compatibility with scaling within minute errors!
We have also computed the static potentials at $\beta=2.2$
for other values of $\lambda$ in the range $0.5\le\lambda<0.8$.
No dependence on $\lambda$ is seen.
These results are a strong indication for a continuum-like behavior of the
static potentials already at the small $\beta$ values that we use.

\section{Conclusions}

We clearly observe string breaking as level crossing phenomenon in the
confinement ``phase'' of the SU(2) Higgs model. The static potentials
show a remarkable scaling behavior under variation of the lattice
spacing by almost a factor two. The search for a method of defining
lines of constant physics in the confinement ``phase'' of the SU(2)
Higgs model has only partially been solved (\fig{f_lcpkl}). Finding
a renormalised observable sensitive to $\lambda$ is still an open and
interesting question.

I thank R. Sommer for his constant support and the
Konrad-Zuse-Zentrum {f\"ur} Informationstechnik Berlin (ZIB)
for granting CPU-resources to this project.


\begin{thebibliography}{9}

\bibitem{schilling} For a review see K. Schilling, these proceedings.

\bibitem{adjpot:su2michael} C. Michael, Nucl. Phys. B (Proc. Suppl.)
26 (1992) 417.

\bibitem{4dhiggs} ALPHA, F. Knechtli and R. Sommer, Phys. Lett. B440
(1998) 345, erratum: Phys. Lett. B454 (1999) 399.

\bibitem{phd} F. Knechtli, Ph.D. thesis, to appear in hep-lat.

\bibitem{phaseshifts:LW} M. L\"uscher and U. Wolff, Nucl. Phys. B339
                         (1990) 222.

\bibitem{pot:r0} R. Sommer, Nucl. Phys. B411 (1994) 839.

\bibitem{Higgs:Montvay1} I. Montvay, Nucl. Phys. B269 (1986) 170.

\bibitem{drummond} I.T. Drummond, Phys. Lett. B434 (1998) 92.

\end{thebibliography}
\end{document}